\documentclass{IEEEtran}
\usepackage{cite}
\usepackage{amsmath,amssymb,amsfonts}
\usepackage{graphicx}
\usepackage{textcomp,nicefrac}
\def\BibTeX{{\rm B\kern-.05em{\sc i\kern-.025em b}\kern-.08em
T\kern-.1667em\lower.7ex\hbox{E}\kern-.125emX}}
\begin{document}
\title{The DUNE-DAQ Application Framework}
\author{Eric L. Flumerfelt for the DUNE Collaboration
\thanks{Paper submitted for review May 16, 2024. Notice: This work was produced by Fermi Research Alliance, LLC under contract No. DEAC02-07CH11359 with the U.S. Department of Energy,Office of Science, Office of High Energy Physics. The United States Government retains and the publisher, by accepting the work for publication, acknowledges that the United States Government retains a non-exclusive, paid-up, irrevocable, world-wide license to publish or reproduce the published form of this work, or allow others to do so, for United States Government purposes. The Department of Energy will provide public access to these results of federally sponsored research in accordance with the DOE Public Access Plan (http://energy.gov/downloads/doe-public-access-plan).}
\thanks{Eric L. Flumerfelt is with the Fermi National Accelerator Laboratory, Batavia, IL 60510 USA.}
}

\maketitle

\begin{abstract}
The Deep Underground Neutrino Experiment (DUNE) is a next-generation neutrino experiment that will probe the properties of these elusive particles with unparalleled precision. It will also act as an observatory for neutrino bursts caused by nearby supernovae, in the event that one occurs while the experiment is in operation. Given these goals, the DUNE trigger and DAQ system must be able to maintain extremely high uptime and provide a path for full readout of the detectors for very long times (up to 100~s). To achieve these ends, we have designed the DUNE DAQ system around a flexible “application framework”, which provides a modular interface for specific tasks while handling the interconnections between them. The application framework collects modules into applications which can then be interacted with as units by the control, configuration and monitoring systems. One of the key features of the framework is its communications abstraction layer, which allows for modules to interact with both internal queues and external network connections with a single transport-agnostic interface. We will report on the architecture and features of the framework.
\end{abstract}


\section{Introduction}
\label{sec:introduction}
\begin{figure}[t]
\centerline{\includegraphics[width=3.5in]{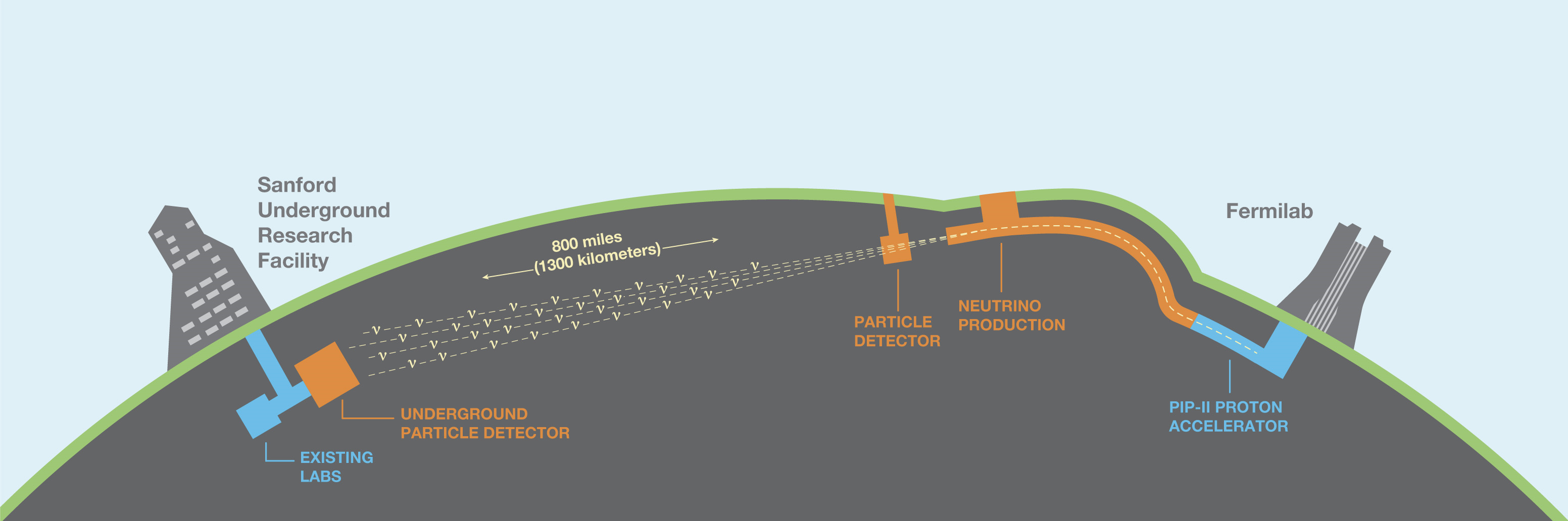}}
\caption{This cutaway illustration shows the path of neutrinos in the Deep Underground Neutrino Experiment. A proton beam is produced in Fermilab’s accelerator complex (improved by the PIP-II project). The beam hits a target, producing a neutrino beam that travels through a particle detector at Fermilab, then through 800 miles (1,300 km) of earth, and finally reaches the far detectors at Sanford Underground Research Facility.}
\label{beamline}
\end{figure}

\IEEEPARstart{T}{he} Deep Underground Neutrino Experiment (DUNE) \cite{dune} is a next-generation neutrino experiment that will probe the properties of these elusive particles with unparalleled precision. It is based in Lead, SD and receives neutrino beam from Fermilab, a 1300~km baseline, as seen in \figurename~\ref{beamline}. 

\begin{figure}[t]
\centerline{\includegraphics[width=3.5in]{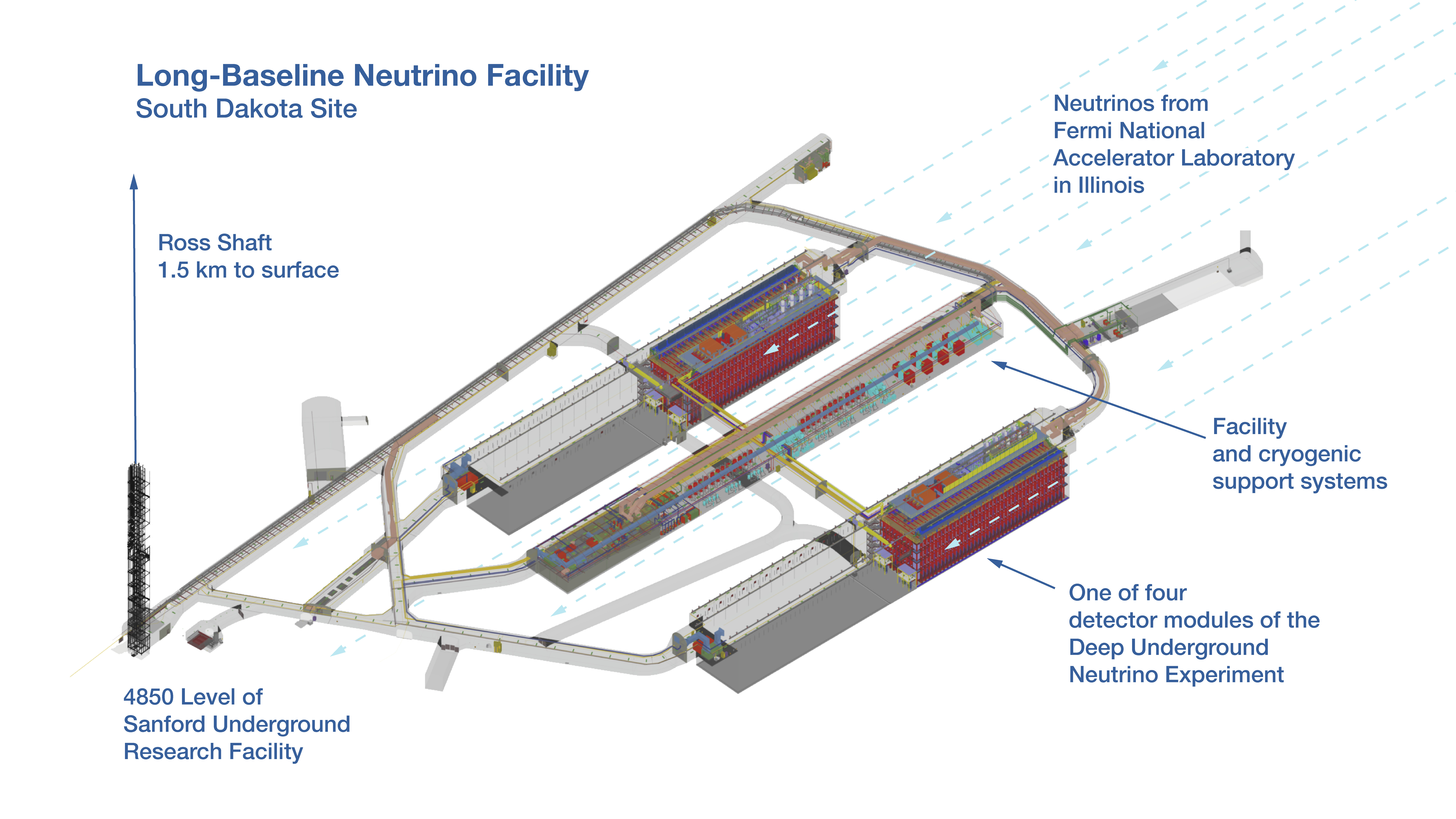}}
\caption{Long-Baseline Neutrino Facility South Dakota Site}
\label{cavern}
\end{figure}

Up to four far detector modules will be constructed, with the first two being liquid-Argon time-projection chambers, \figurename~\ref{cavern}.  It will also act as an observatory for neutrino bursts caused by nearby supernovae, in the event that one occurs while the experiment is in operation.

Given these goals, the DUNE trigger and DAQ system \cite{dunedaq} must be able to maintain extremely high up-time and provide a path for full readout of the detectors for very long times (up to 100~s). To achieve these ends, we have designed the DUNE DAQ system around a flexible “application framework”, which provides a modular interface for specific tasks while handling the interconnections between them.

During the first DUNE 35~T and initial ProtoDUNE test runs, the \textit{artdaq} DAQ framework from Fermilab was used \cite{artdaq}. The up-time requirement for DUNE DAQ is 99.5\%, leading to a requirement that the DAQ be able to handle run-time changes of the components selected for readout. It was found, however, that \textit{artdaq} lacked sufficient flexibility to provide this functionality. Additionally, the \textit{artdaq} system layout did not provide hooks for some of the data processing steps needed at the readout level, resulting in less-maintainable plugin implementations. 

The DUNE DAQ application framework was designed to meet the requirements of the DUNE. It provides a common set of APIs and utilities that allow for implementation of the DAQ logic in a consistent and maintainable way, while also providing upgradability through the use of dynamically-loaded plugin implementations for well-defined application interfaces.

\section{DUNE DAQ Overview}
\begin{figure}[t]
\centerline{\includegraphics[width=3.5in]{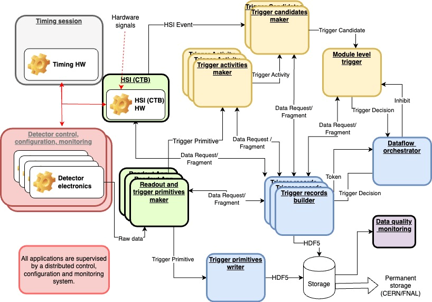}}
\caption{Current organization of the DUNE DAQ software applications. Messaging connections between applications are shown along with their data types, though it is important to note that DAQ modules within the applications are sending and receiving messages, not the applications themselves.}
\label{dunedaq}
\end{figure}

The DUNE DAQ consists of a number of application components which communicate with one another as in \figurename~\ref{dunedaq}. Each application handles a specific set of DAQ functionality using a number of "DAQ Modules" specific to that application. 

\begin{figure}[t]
\centerline{\includegraphics[width=3.5in]{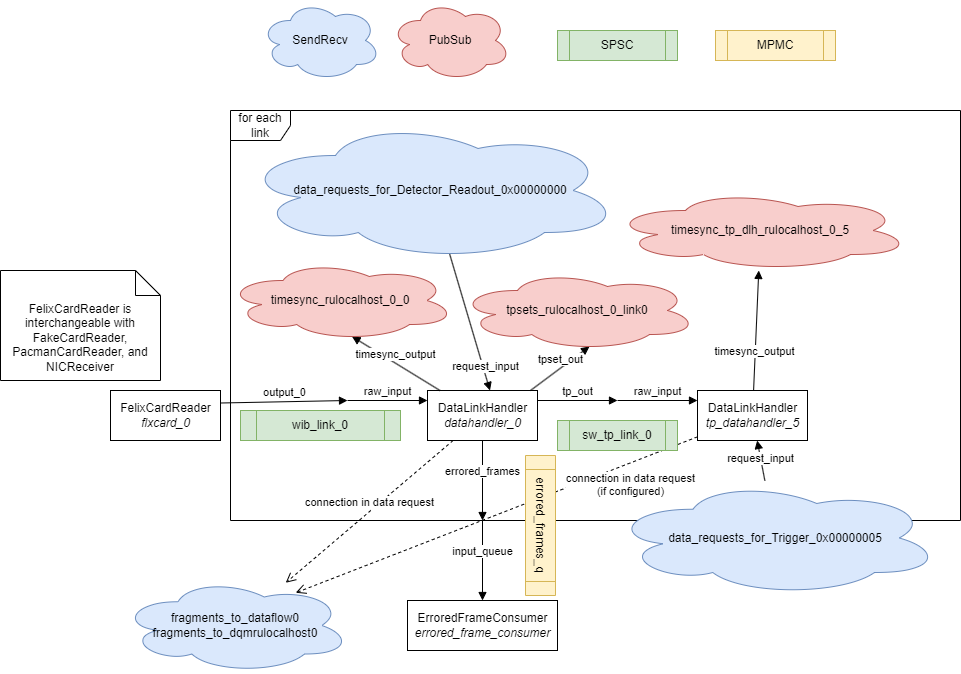}}
\caption{Diagram of the readout application used in the DUNE DAQ. Modules are represented by white boxes, and network connections and queues are also shown. Queues are always strictly internal to the application, whereas network connections are always single-sided from the application's perspective.}
\label{readout}
\end{figure}

\begin{enumerate}
\item Readout applications receive and buffer data from detector electronics, and send processed waveform data (called Trigger Primitives) to the trigger applications. They listen for incoming data requests from the builder application. The internal structure of a readout application is depicted in \figurename~\ref{readout}.
\item The Hardware Signals Interface (HSI) application interfaces with the timing hardware and the Central Trigger Board (CTB) to produce HSI Event messages for periodic and beam event triggers.
\item The trigger applications create trigger decisions based on configurable algorithms using the trigger primitives sent from the readout. The trigger primitives are broadcast to all applications which have registered themselves as receivers, so the number of trigger applications is flexible and dependent on the number of desired trigger algorithms. 
\item The trigger record builder applications receive trigger decisions, request data from the readout and trigger applications, and build Trigger Records which are saved to disk. 
\item The Dataflow Orchestrator (DFO) application arbitrates the distribution of trigger decisions to builder applications based on current run-time conditions, and protects the builder applications from becoming overwhelmed by large trigger records, inhibiting the production of new trigger decisions if necessary. 
\item An additional Trigger primitive writer application may be used to save the Trigger Primitive stream to disk; these files can then be used for future trigger algorithm development.
\end{enumerate}

\section{Application Framework Structure}
\begin{figure}[t]
\centerline{\includegraphics[width=3.5in]{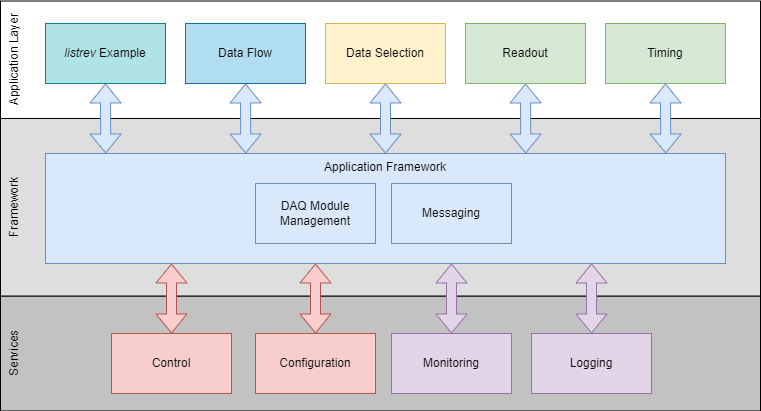}}
\caption{Logical stack diagram of the application framework. Core services are exposed to the framework through interfaces, with plugin-based implementations to allow for upgrading functionality as technologies evolve during the life-time of the experiment. The framework's core functionality includes the definition of applications, the management of DAQ modules, and the messaging API used by the modules. Applications of various types are implemented on top of the framework to perform the data acquisition logic required by the experiment.}
\label{stack}
\end{figure}

\begin{figure}[t]
\centerline{\includegraphics[width=3.5in]{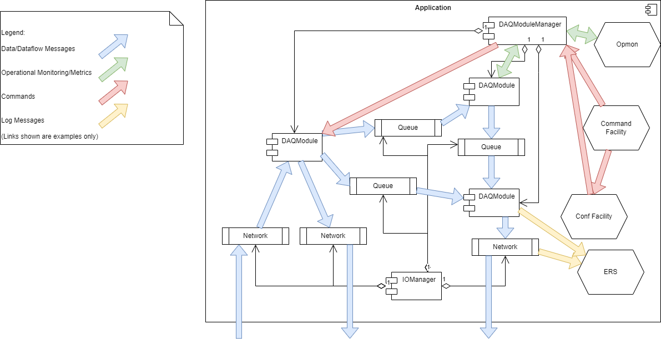}}
\caption{Diagram of a generic application based on the framework, showing queues and external network connections. Several facilities are available to the application and exposed to DAQ modules through framework interfaces.}
\label{application}
\end{figure}

\begin{figure}[t]
\centerline{\includegraphics[width=3.5in]{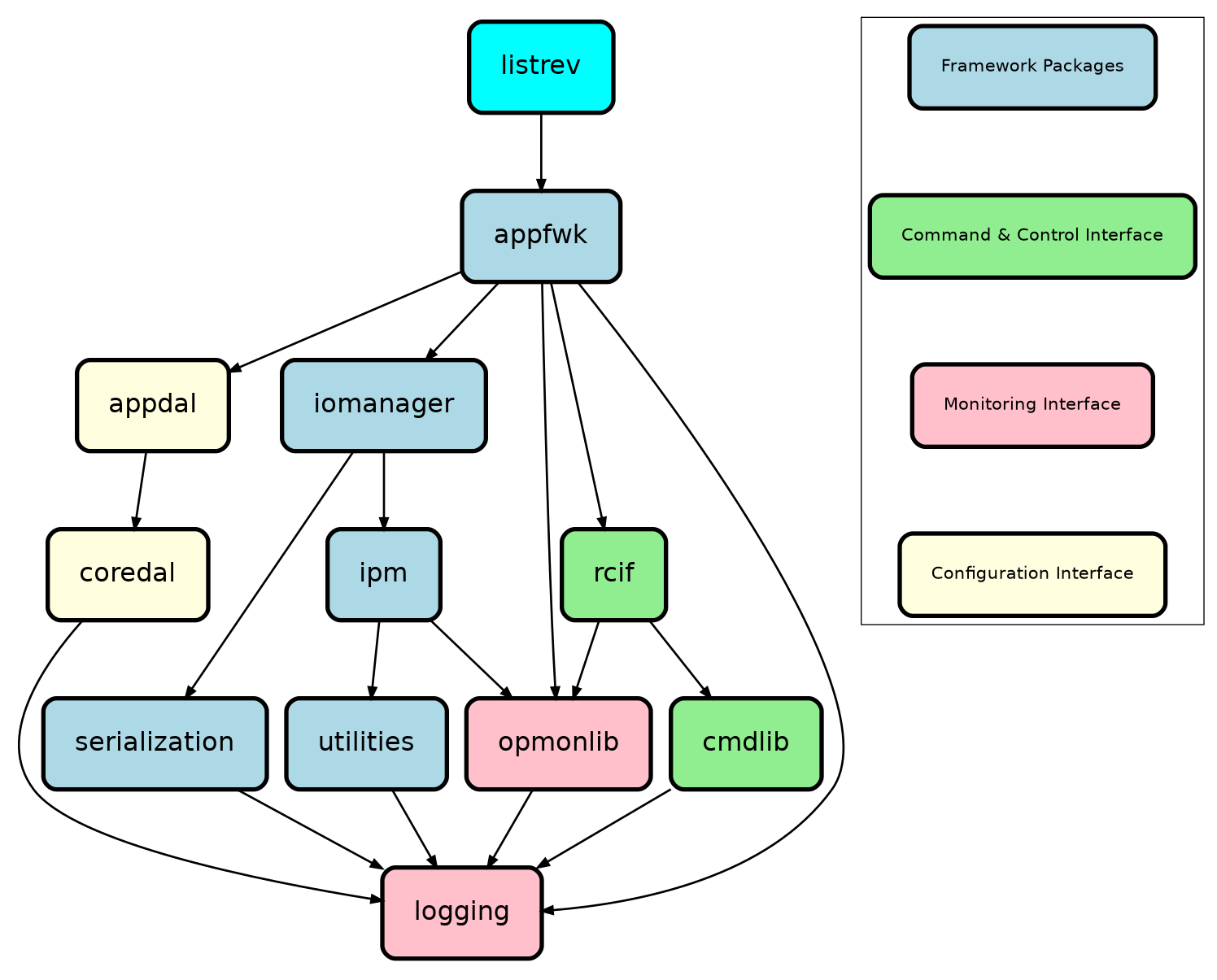}}
\caption{Software packages comprising the application framework. The listrev package is shown as an example user of the framework, while several of the packages containing service interfaces and implementations are also shown as dependencies of the framework.}
\label{packages}
\end{figure}

The DUNE DAQ Application Framework consists of several software packages which implement various sub-components and expose interfaces to the central framework functionality, as shown in \figurename~\ref{stack}, and the \textit{appfwk} \cite{appfwk} package which provides the main interfaces used by the implementation modules. As previously stated, applications are composed of a number of DAQ modules which implement the functionality for each application type, \figurename~\ref{application}. Services used by the application framework abstract their implementations behind consistent interfaces so that those implementations can be changed at-will, increasing the longevity and maintainability of the framework.
\figurename~\ref{packages} shows the structure of the application framework packages, as well as including one of the packages which uses the interfaces provided by the application framework (\textit{listrev} \cite{listrev} is the application framework test package). 
\begin{enumerate}
\item The \textit{iomanager} \cite{iomanager} package provides the messaging interface used by the modules, whereas the DAQ module management facilities and API are implemented in the \textit{appfwk} package itself. 
\item The \textit{ipm} \cite{ipm} and \textit{serialization} \cite{serialization} packages are used for transmitting data between applications.
\item \textit{utilities} \cite{utilities} contains a number of tools that are used throughout the framework (including DNS resolution, a generic base class for all application framework objects, and threading tools).
\item \textit{rcif} \cite{rcif} provides the interface to the run control system, implemented in \textit{cmdlib} \cite{cmdlib}. 
\item \textit{appdal} \cite{appdal} and \textit{coredal} \cite{coredal} provide the configuration interface and schema, as well as recipes for generating the most commonly-used applications and their messaging connections at run-time, using module configurations stored in the configuration database.
\item The \textit{opmonlib} \cite{opmonlib} package contains the interface and implementation of the DAQ monitoring and metric collection code.
\item \textit{logging} \cite{logging} takes care of log message archiving and provides a fast, run-time configurable interface for debug messages. 
\end{enumerate}

\subsection{Service Interfaces}
The Application Framework provides a plug-in interface for these external interfaces, allowing for both custom and commercial off-the-shelf implementations to be used.

The Control interface keeps track of application state and validates incoming commands. Command transmission between applications can use a REST or gRPC-based \cite{grpc} communication implementation. The Run Control system is a separate set of applications which are outside of the application framework.
When a valid command is received, the application distributes it to its DAQ modules via an "action plan", a pre-configured series of parallel steps. Each step of an action plan is an "action", an aliased method within the DAQ module. A single module may register multiple actions to be executed within a single action plan, in order to define "pre" and "post" command steps.

Configuration information is provided to the application at startup using the OKS system adapted from ATLAS \cite{atlas}. The configuration system uses an object-based model and retrieves configuration data based on object class and unique names. Along with the logging interface, the configuration interface is one of the basic components of the system which must be configured via the application command-line; all other service interfaces and application behavior are defined through the configuration.

Each "monitorable" element of the system is given a unique name which it can use to push metrics to the monitoring interface, which in turn publishes them using a plugin-based output facility. Additionally, the monitoring interface periodically polls for metrics in a top-down manner, ensuring that all monitored elements regularly report their current values. Current output facility implementations include JSON files and a protobuf \cite{protobuf} format via a Kafka \cite{kafka} broker. If distributed via Kafka, the data are stored in an InfluxDB \cite{influxdb} database and plots are automatically construted via DUNE-developed Grafana \cite{grafana} dashboards.

Log messages are handled using the ERS package \cite{ers}, also adapted from ATLAS \cite{atlas-ers}. ERS provides for well-structured messages which can be collected in a back-end database (currently implemented in Kafka) for further processing. Additionally, Fermilab’s TRACE \cite{trace} package provides high-speed low-level debug printouts which are assigned a name and a level, and these can be enabled or disabled at run-time. The performance of TRACE has been measured at \textless 1~\textmu s for enabled messages and O(1~ns) for disabled messages.

\subsection{DAQ Module Interface}
The goal of the Application Framework is to simplify as much as possible the interactions with other systems, so that DAQ modules can be written in a self-contained, task-oriented manner. It consolidates interfaces for control, configuration, monitoring, and logging; these systems interact with applications, allowing them to ignore the internal complexity of the modules and their interconnections. Users are provided with access to data received from the configuration system and the interface used to aggregate monitoring data. The framework marshals and delivers commands to modules when they are received by the applications.

The modules register their interest in certain commands, so when a command is received, the application can intelligently deliver it to just the modules that have a corresponding action. The framework also provides an abstraction layer for connections, allowing modules to send data to an endpoint identified by a name without regard as to whether the module owning the endpoint is local to the application or running elsewhere.

Apart from a single required \texttt{init} method, the developer is free to implement whatever methods and internal threading is needed to accomplish the task of the DAQ module. The encapsulation provided by the messaging system means that the module is completely independent of other modules, though it is recommended for modules to use the callback system provided by the messaging API to schedule work on message arrival.

\subsection{Messaging API}
One of the core functionalities of the Application Framework is its messaging API. Each message channel consists of a unidirectional, strongly-typed connection (i.e. each connection only handles one type of message data), with both network and queue-based implementations for inter- and intra-process messaging, respectively. Connections are indexed by name and data type (converted to a string by macros defined in the \textit{serialization} package and called at data type definition), and can be retrieved from an external connectivity service using pattern matching on the name. Supported topologies include point-to-point, broadcast, and many-to-one connections. Endpoints are classified as being "bind"-type or "connect"-type, based on whether the endpoint should be at a well-known or ephemeral location for the given topology. For example, in the many-to-one case, the receiver is the "bind"-type endpoint, and all senders are "connect" type. Meanwhile, for the broadcast pattern, the sender is the "bind"-type and any receivers are "connect" type (this particular topology also allows for zero receivers).

The messaging API exposed to the DAQ modules is agnostic for whether messages are internal to an application or sent over the network, allowing for greater flexibility in how modules are organized into applications. This also allows for custom applications implementing large portions of the DAQ logic to be created for smaller test scenarios. Conversely, if a particular application has a large processing load, it can be split up to take advantage of available computing resources.

Network transport and serialization use plug-in interfaces, currently implemented via ZMQ \cite{zmq} (in \textit{ipm} and MsgPack \cite{msgpack}). ZeroMQ "push/pull" and "pub/sub" socket types are used to implement the required connection topologies. Several ZeroMQ socket options are used to guarantee that messages are only accepted if the remote endpoints are connected, including \texttt{sndtimeo} and \texttt{rcvtimeo} set to 0 and \texttt{immediate} set on the sender socket. These options allow for the application to quickly detect when remote endpoints are not functional and attempt to reconnect.

\subsection{Threading and Utilities}
\begin{figure}[t]
\centerline{\includegraphics[width=3.5in]{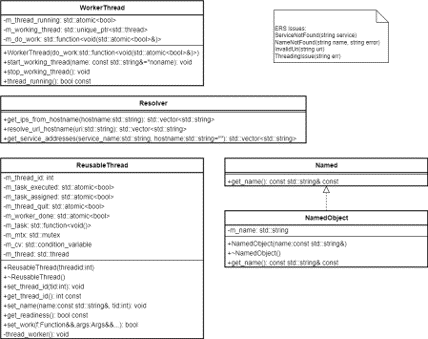}}
\caption{UML Class diagrams of the application framework utilities.}
\label{utilities}
\end{figure}

Several utility classes are provided by the Application Framework, diagrammed in \figurename~\ref{utilities}. Most notable are the WorkerThread and ReusableThread interfaces, which provide long-lived tasks and a thread pool implementation, respectively. These threading helper classes help DAQ module developers to implement their logic in a consistent way, improving system maintainability and allowing for the framework to manage thread lifetime and exception handling.

The framework provides a naming hierarchy which is used for monitoring and logging, and the \textit{utilities} package provides the \texttt{NamedObject} interface which is the base class for all application framework classes.

DNS resolution utilities provided by the \texttt{Resolver} class are used in the network messaging implementation and have the ability to interface with the Kubernetes DNS \cite{k8s} for service-based endpoint resolution.

\section{Testing and Examples}

In order to ensure that the application framework meets the requirements of the DUNE DAQ, each component of the system is tested in both individual unit tests and through a series of integration tests of increasing complexity, cumulating in tests that exercise a significant portion of the DAQ itself, including real detector hardware.

Each package within the application framework has a suite of unit tests, and the \textit{listrev} example was created to exercise the functionality of the framework as whole as part of the integration testing of the DUNE DAQ. Its operation is as follows:
\begin{enumerate}
\item A "ReversedListValidator" module broadcasts a request to create a new list of integers. 
\item Each RandomDataListGenerator uses the information in the create message to product a list of integers. 
\item The ReversedListValidator requests a list identified by an iteration index from the RandomDataListGenerators and the ListReverser. 
\item The ListReverser, upon receiving a request for a new list, in turn requests the list from the RandomDataListGenerator, inverts the order of the list received, and sends it to the ReversedListValidator.
\item Once the ReversedListValidator has received the relevant pairs of data, it can check that the list was indeed reversed and that it matches the pattern specified in the create message, seeded by the iteration index.
\end{enumerate}
This example showcases many different aspects of the application framework's functionality, including DAQ module creation and managment, several messaging topologies and patterns (create messages use broadcasts, and the request of one data type resulting in a data response of a different data type), and it uses the control, monitoring, and configuration interfaces. This allows for a discrete test of the application framework without relying on detector hardware in well-controlled conditions.

\begin{figure}[t]
\centerline{\includegraphics[width=3.5in]{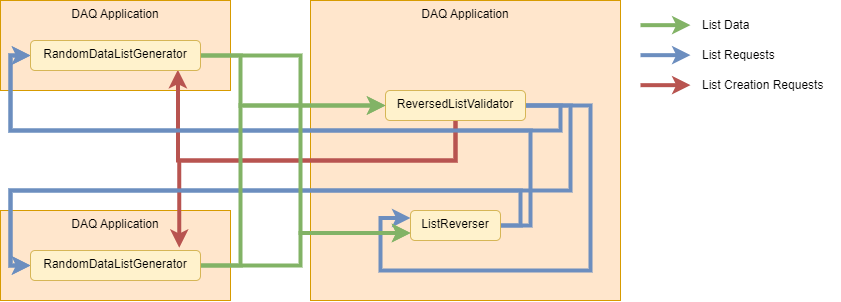}}
\caption{Communication paths within the listrev example package. Several iterations of the example are run as part of the application framework integration test suite, to ensure that queues and network connections are properly working in the required connectivity patterns, and that the DAQ modules remain functional regardless of which application they are located in or what other DAQ modules are also in that application. The configuration shown has two generators in separate applications and one reverser and validator in a third application.}
\label{listrev}
\end{figure}

\section{Implementations of the Framework}
The DUNE DAQ Application Framework has been used as the basis for the DAQ system for ProtoDUNE horizontal and vertical drift, as well as the ICEBERG prototype detector at Fermilab and the TOAD Near Detector prototype.
The system has evolved with each iteration, with different DAQ modules being used in different configurations. Readout has transitioned from Proto-WIBs to DUNE-WIBs and from FELIX-based readout \cite{felix} to WIBEth Ethernet-based readout \cite{wibeth}. This has shown the framework's ability to interface with different readout hardware with a reasonable development period for the DAQ modules specific to each readout type.
The DUNE-DAQ Application Framework is being deployed at ProtoDUNE II for data-taking activities in 2024, with the horizontal-drift TPC currently being made ready for operations.

\section{Conclusions}
The application framework has been deployed at the ProtoDUNE test detector at CERN, as well as other test detectors, and we have used the test runs to refine and improve the framework. The framework is in the final stages of development, with reviews being performed to finalize the functionality and close out development of new features. As it is a central part of the DUNE DAQ, any API changes results in a large amount of effort to reconcile existing module implementations with the change, so it is imperative to have the framework in a very stable state to allow development to reach a conclusion in other parts of the system.

We have demonstrated that the framework's plugin-based architecture allows for easy extension to additional readout types, an essential feature for the DUNE which will require readout from at least three and up to six different detector technologies simultaneously. We believe that the framework may prove useful as the basis for the DAQ suites of other high-energy physics experiments in the future.


\begin{thebibliography}{00}
\bibitem{dune} \textit{Dune homepage}. [Online]. Available: \underline {https://dunescience.org}. Accessed on: May 7, 2024.
\bibitem{dunedaq} A. Abed Abud, C. Batchelor, K. Biery et al., ``Trigger and data acquisition (tdaq) system design,'’ DUNE DAQ Project, Tech. Rep., Jan. 2023. [Online]. Available: \underline{https://edms.cern.ch/document/2812882}. Accessed on: May 7, 2024.
\bibitem{artdaq} K. Biery, E. Flumerfelt, J. Freeman, W. Ketchum, G. Lukhanin, and R. Rechenmacher, ``Artdaq: DAQ software development made simple,'' \textit{Journal of Physics: Conference Series}, vol. 898, no. 3, pp. 032013, Oct. 2017, DOI.
 10.1088/1742-6596/898/3/032013.
\bibitem{appfwk} \textit{DUNE-DAQ/appfwk}. [Online]. Available: \underline{https://github.com/DUNE-DAQ/appfwk}. Accessed on: May 7, 2024
\bibitem{listrev} \textit{DUNE-DAQ/listrev}. [Online]. Available: \underline{https://github.com/DUNE-DAQ/listrev}. Accessed on: May 7, 2024
\bibitem{iomanager} \textit{DUNE-DAQ/iomanager}. [Online]. Available: \underline{https://github.com/DUNE-DAQ/iomanager}. Accessed on: May 7, 2024
\bibitem{ipm} \textit{DUNE-DAQ/ipm}. [Online]. Available: \underline{https://github.com/DUNE-DAQ/ipm}. Accessed on: May 7, 2024
\bibitem{serialization} \textit{DUNE-DAQ/serialization}. [Online]. Available: \underline{https://github.com/DUNE-DAQ/serialization}. Accessed on: May 7, 2024
\bibitem{utilities} \textit{DUNE-DAQ/utilities}. [Online]. Available: \underline{https://github.com/DUNE-DAQ/utilities}. Accessed on: May 7, 2024
\bibitem{rcif} \textit{DUNE-DAQ/rcif}. [Online]. Available: \underline{https://github.com/DUNE-DAQ/rcif}. Accessed on: May 7, 2024
\bibitem{cmdlib} \textit{DUNE-DAQ/cmdlib}. [Online]. Available: \underline{https://github.com/DUNE-DAQ/cmdlib}. Accessed on: May 7, 2024
\bibitem{appdal} \textit{DUNE-DAQ/appdal}. [Online]. Available: \underline{https://github.com/DUNE-DAQ/appdal}. Accessed on: May 7, 2024
\bibitem{coredal} \textit{DUNE-DAQ/coredal}. [Online]. Available: \underline{https://github.com/DUNE-DAQ/coredal}. Accessed on: May 7, 2024
\bibitem{opmonlib} \textit{DUNE-DAQ/opmonlib}. [Online]. Available: \underline{https://github.com/DUNE-DAQ/opmonlib}. Accessed on: May 7, 2024
\bibitem{logging} \textit{DUNE-DAQ/logging}. [Online]. Available: \underline{https://github.com/DUNE-DAQ/logging}. Accessed on: May 7, 2024
\bibitem{grpc} \textit{gRPC}. [Online]. Available: \underline{https://grpc.io/}. Accessed on: May 7, 2024
\bibitem{atlas} R. Jones, L. Mapelli, Y. Ryabov, and Soloviev, I., ``The OKS persistent in-memory object manager,'' \textit{IEEE Trans. Nuclear Science}, vol. 45, no. 4, pp. 1958-1964, Aug. 1998, DOI. 10.1109/23.710971
\bibitem{protobuf} \textit{Protocol Buffers Documentation}. [Online]. Available: \underline{https://protobuf.dev/}. Accessed on: May 7, 2024
\bibitem{kafka} \textit{Apache Kafka}. [Online]. Available: \underline{https://kafka.apache.org/}. Accessed on: May 7, 2024
\bibitem{influxdb} \textit{InfluxDB Overview}. [Online]. Available: \underline{https://www.influxdata.com/products/influxdb-overview/}. Accessed on: May 7, 2024
\bibitem{grafana} \textit{Grafana: The open observability platform}. [Online]. Available: \underline{https://grafana.com/}. Accessed on: May 7, 2024
\bibitem{ers} \textit{DUNE-DAQ/ers}. [Online]. Available: \underline{https://github.com/DUNE-DAQ/ers}. Accessed on: May 7, 2024
\bibitem{atlas-ers} S. Kolos,A. Kazarov, and L. Papaevgeniou, ``The Error Reporting in the ATLAS TDAQ System,'' \textit{Journal of Physics: Conference Series}, vol. 608, no. 1, pp. 012004, Apr. 2015, DOI. 10.1088/1742-6596/608/1/012004
\bibitem{trace} S. Foulkes, and R. Rechenmacher, ``TRACE - A System Wide Diagnostic Tool,'' \textit{2007 IEEE-NPSS Real-Time Conference}, 2007, DOI. 10.1109/RTC.2007.4382742
\bibitem{zmq} \textit{ZeroMQ: An open-source universal messaging library}. [Online]. Available: \underline{https://zeromq.org/}. Accessed on: May 7, 2024
\bibitem{msgpack} \textit{MessagePack: It's like JSON, but fast and small}. [Online]. Available: \underline{https://msgpack.org/index.html}. Accessed on: May 7, 2024
\bibitem{k8s} \textit{DNS for Services and Pods | Kubernetes}. [Online]. Available: \underline{https://kubernetes.io/docs/concepts/services-networking/dns-pod-service/}. Accessed on: May 7, 2024
\bibitem{felix} Soo Ryu on behalf of the ATLAS TDAQ Collaboration, ``FELIX: The new detector readout system for the ATLAS experiment,'' \textit{Journal of Physics: Conference Series}, vol. 898, no. 3, pp. 032057, Oct. 2017, DOI. 10.1088/1742-6596/898/3/032057
\bibitem{wibeth} R. Sipos for the DUNE collaboration, ``The Ethernet readout of the DUNE DAQ system,'' \textit{2024 IEEE-NPSS Real-Time Conference}, Apr. 2024, to be published
\end{thebibliography}
\end{document}